\documentclass[prd,showpacs,superscriptaddress]{revtex4}
\usepackage{amsmath}
\usepackage{amssymb}
\usepackage{color}
\usepackage{graphicx}

\begin{document}

\title{Understanding the cosmic ray positron flux}

\author{Paolo Lipari}
\email{paolo.lipari@roma1.infn.it}
\affiliation{INFN, Sezione Roma ``Sapienza'',
 piazzale A.Moro 2, 00185 Roma, Italy}


\begin{abstract}
Recent precision measurements of the flux of cosmic ray positrons
by the Alpha Magnetic Spectrometer show that the spectrum has
a marked softening feature for energies close to one TeV.
A possible interpretation of this result is that the
observed  feature  measures the maximum  energy  of
a new hard source of positrons perhaps associated to dark matter
self--annihilation or decay,
or to positron accelerators.  A gradual  hardening of the
positron flux  centered at  $E \simeq 25$~GeV can also
be understood as the signature of the transition where
the new source 
overtakes the conventional component  due to  secondary production.
This interpretation is simple and attractive, but it is not unique.
The alternative possibility,  that the positron flux
is entirely of secondary origin,  remains viable.
In such a scenario the spectral softening observed
by AMS for positrons  is  generated by energy loss effects,
and  a feature of similar, but not
identical structure should be also  visible in the $e^-$ spectrum.
Spectral features similar to both  the hardening
and softening of the positron flux  are in fact 
observed for electrons and call for a consistent explanation.
Precision measurements of the $e^+$ and $e^-$ spectra in the
TeV and multi--TeV energy range are crucial to clarify  the problem.
\end{abstract}

\maketitle

\section{Introduction}
\label{sec:intro} 
The cosmic ray (CR) positrons flux is
of great importance for High Energy Astrophysics because it can be
a probe to investigate the possible existence of
Galactic dark matter in the form of Weakly Interacting Massive Particles,
and of astrophysical antimatter accelerators. 
The shape of the $e^+$
spectrum gives also very valuable information
to determine the properties of propagation of CR particles in the Milky Way.

Recently the AMS Collaboration 
has released new data on the positron spectrum \cite{Aguilar:2019owu}
that extend the measurements to a maximum energy of 1~TeV.
In this paper we study the shape of the
$e^+$ spectrum, compare it to the spectra of
other particles (in particular $e^-$ and $\overline{p}$)
and discuss  possible interpretations of the observations.

\begin{figure}[bt]
\begin{center}
\includegraphics[width=14.0cm]{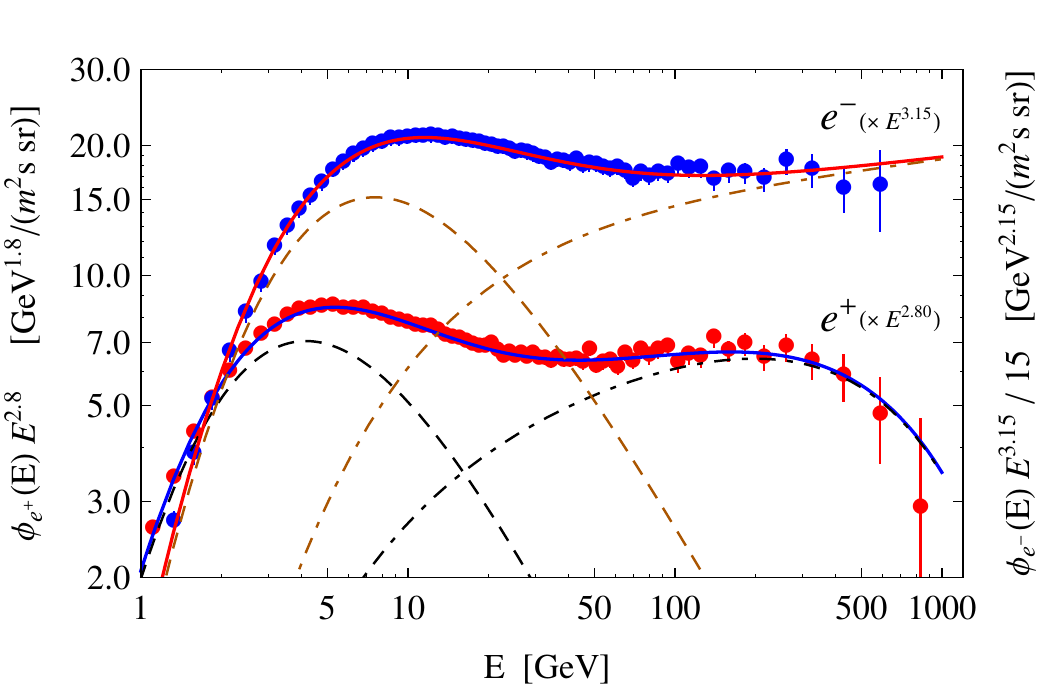}
\end{center}
\caption {\footnotesize
 Spectra of CR positrons \cite{Aguilar:2019owu}
 and electrons \cite{ams-electrons-positrons-2014} measured
 by AMS02.
 The fluxes are plotted as a function of energy
 in the form $E^{2.8}~\phi(E)$ for positrons and
 $E^{3.15}~\phi(E)$ for electrons.
 The lines are two components calculated 
 by the AMS Collaboration in \cite{Aguilar:2019owu} for positrons
 and in \cite{Lipari:2018usj} for electrons.
\label{fig:flux_components}}
\end{figure}

The new AMS  data on positrons are shown in Fig.~\ref{fig:flux_components}
(plotted together with data on electrons) and in Fig.~\ref{fig:fit_all}
[together with measurements of the fluxes  of
  $\overline{p}$ and  and ($e^- + e^+$)].
The AMS Collaboration in \cite{Aguilar:2019owu} has fitted the
positron data using the functional form
\begin{equation}
 \phi_{e^+}(E) = C_1 ~\left (\frac{E}{E_0} \right)^{-\gamma_1}
 + C_s~\left (\frac{E}{E_0} \right)^{-\gamma_s} ~e^{-E/E_s}
\label{eq:ams-fit}
\end{equation}
(with $E_0$ an  arbitrary energy scale) modified by solar modulations
described by the force field approximation (FFA) \cite{Gleeson:1968zza}.
With this assumption the directly observable flux takes the form
\begin{equation}
\phi_{e^+}^{\rm obs} (E) = \phi_{e^+}(E + \varphi) \; \frac{E^2}{(E+ \varphi)^2} ~,
\label{eq:modulations}
\end{equation}
with $\varphi$ a time
dependent parameter with the dimension of energy.

In the expression of Eq.~(\ref{eq:ams-fit})
the $e^+$ spectrum is described as the sum of two distinct components
(that are also shown in Fig.~\ref{fig:flux_components} together with
their sum).
The first component has a simple power law form with
(best fit) spectral index \cite{f:spectral-index}
$\gamma_1 = 4.07$ and dominates at low energy.
The second component is a harder power law
with spectral index $\gamma_s \simeq 2.58$,
and becomes dominant for $E \gtrsim 20$~GeV.
At high energy this second component (and therefore the
observable flux) has also a marked softening
feature that is described as an exponential cutoff
\cite{f:suppression-remark}.
The parameter $E_s$ is determined with a large error $E_s = 810^{+310}_{-180}$~GeV.

The functional form of Eq.~(\ref{eq:ams-fit}) 
(with identical modeling of solar modulation effects,
but without the high energy cutoff) had already been used
in \cite{Lipari:2018usj} to fit the data on the positron and electron 
spectra previously released by
PAMELA \cite{pamela-positrons-2013,pamela-electrons-2011}
and AMS \cite{ams-electrons-positrons-2014},
with results that for the positrons are
in very good agreement with those presented in \cite{Aguilar:2019owu} 
(the best fit spectral indices for the two components
$\gamma_1 \simeq 3.99$ and $\gamma_s \simeq 2.66$).
The larger exposure in the new AMS paper extends the measurements
of the positron flux to higher energy and requires the introduction of
a softening at high energy to describe the data.
The observation of this spectral suppression is the
new, important result in the AMS paper.

The AMS Collaboration argues in \cite{Aguilar:2019owu}
that the new data on the $e^+$ spectrum require the existence
of a new positron source
in addition to the conventional mechanism
of secondary production, where the positrons are created 
in the inelastic collisions of CR protons
and nuclei with interstellar gas. 
In this interpretation the contribution of the new, non--conventional source 
dominates the $e^+$ flux at high energy, and 
corresponds to the second, harder component
in Eq.~(\ref{eq:ams-fit}). Its origin 
could be associated to the self--annihilation or decay of dark matter
particles, or to astrophysical positron accelerator.
Such a conclusion is of great importance and should be
carefully scrutinized.

Schematically, there are three observations (and three
arguments associated to the observations) that give support to the
hypothesis of the existence of a non--conventional component in the
positron spectrum. \\
(A) In a broad energy interval ($E \simeq 30$--400~GeV) the $e^+$ flux
has a spectral shape % (a power law with index $\gamma \approx 2.7$)
that is significantly harder than the ``conventional'' prediction
for positrons generated by the secondary production mechanism. \\
(B) The flux exhibits a gradual hardening around the
energy $E_h \simeq 25$~GeV.
This feature, in the fit of Eq.~(\ref{eq:ams-fit})
corresponds to the transition where the hard
component emerges overtaking the soft one. \\
(C) The flux has a marked softening or cutoff at high energy,
that on the AMS fit is described with an exponential factor
$e^{-E/E_s}$ with $E_s \sim 800$~GeV.
The existence of a cutoff in the positron
source spectrum is predicted
in models where the particles
are generated by dark matter
[where the spectrum has a maximum energy at $E = m_\chi$
(for annihilation) or $m_\chi/2$ (for decay),
with $m_\chi$ the mass of the DM particle]
and also in models where positron accelerators have a
sharply defined maximum energy.

In the following we will review the three arguments listed above,
and discuss if the observations can be reconciled with
the hypothesis that the positron flux is entirely of secondary
production origin.

\section{Spectral shape}
\label{sec:shape} 
The commonly accepted (``standard'') prediction
for the shape of the positron flux in the
range $E \simeq 30$--400~GeV is a power law
%form $\phi_{e^+} (E) \propto E^{-\gamma}$
with a spectral index $\gamma_{e^+} \simeq 3.4$--3.7.
This prediction \cite{Moskalenko:1997gh}
was already in existence in 2008
when PAMELA \cite{Adriani:2008zr} reported a much
harder spectrum, and immediately the new data were 
interpreted as the indication
for the existence a non--conventional, hard source of positrons.

The PAMELA observations have been later confirmed
by FERMI \cite{fermi-positrons} and
with higher precision by AMS \cite{ams-electrons-positrons-2014}.
The critical (and still controversial) point
remains the prediction of the shape of the positron flux
when the conventional mechanism of secondary production is dominant.

New predictions of the positron flux, 
still based on the same set of methods and assumptions as the
previous ones, but using 
as input the most recent (and more precise)
data on the spectra of protons and nuclei,
have been recently calculated \cite{Orlando:2017tde,Evoli:2016xgn}.
These calculations continue to predict a positron flux 
much softer than the observations, and support
the idea that a new harder positron source is necessary. 
However a few authors
\cite{Cowsik:2013woa,Blum:2013zsa,Ahlen:2014ica,Lipari:2016vqk}
have discussed the possibility
that the standard prediction is incorrect, and that
the CR positrons can be entirely of secondary origin.

In the following we will schematically outline the fundamental steps
that form the ``standard calculation'' of the positron flux.
The general the relation between the flux $\phi_j(E)$
(observed at the position of
the solar system in the Galaxy) for particle
of type $j$ [$j \in \{p, e^-, e^+, \overline{p}, \ldots\}$]
and its source spectrum $Q_j(E)$
(that is the rate of particles 
injected in interstellar space by all Galactic sources)
can be written in the form:
\begin{equation}
 \phi_{j} (E)
 = \frac{c}{4 \, \pi} ~ 
 Q_j(E)~\frac{\tau_j (E)} {V}
\label{eq:flux-general}
\end{equation}
In this equation the factor $c/(4 \, \pi)$
transforms a particle density into
an isotropic flux (assuming $\beta \simeq 1$),
$V$ is an effective Galactic confinement volume for CR
particles and $\tau_j(E)$ is a quantity with the dimension of time.
Without loss of generality one can
take $V$ as independent from energy and particle type with all
dependence contained into the factor $\tau_j (E)$.
For a realistic choice of the volume $V$,
the quantity $\tau_j (E)$ has the physical meaning of the residence
time in the Galaxy for particles of type $j$ and energy $E$.

There are two main ``sinks'' for relativistic particles
in interstellar space that can balance the injection of the sources:
escape and energy losses \cite{f:interactions}.
Energy losses are only significant for $e^\mp$, so 
for protons, anti--protons and nuclei, the 
characteristic time $\tau_j (E)$ can be identified with the escape time
$T_{\rm esc}$. Since the
containement of CR particles in the Galaxy
is of magnetic nature, the escape time
(for particle with $\beta \simeq 1$) is only a function
of the particle rigidity $p/q$, and in most realistic cases
on the absolute value of the rigidity.
%(in particular if the structure
%of the magnetic field is such that the propagation or the CR particles
%is diffusive)
A common way to parametrize the rigidity dependence
of the escape time is a power law \cite{f:rigidity} of exponent $\delta$
\begin{equation}
 T_{\rm esc} (E) = T_0 \; E^{-\delta} ~.
\label{eq:t-escape}
\end{equation}

For electrons and positrons 
radiative energy losses (that grow $\propto E^2/m^4$) can be important
\cite{f:electrons-positrons},
and one has to take into account
for the loss time (that is the time for an $e^\mp$ to lose half of its energy):
\begin{equation}
 T_{\rm loss}(E) = \frac{E}{\langle |dE/dt|_e \rangle } \simeq
 621.6 ~
\left [ \frac{0.5~{\rm eV}\;{\rm cm}^{-3}}
 {\left \langle \rho_B + \rho_\gamma^* \right \rangle}
\right ]
\left [ \frac{\rm GeV}{E} \right ]~
 ~{\rm Myr}~.
\label{eq:t-loss}
\end{equation}
where $\rho_B$ and $\rho_\gamma$ are the energy densities in magnetic field
and radiation averaged in the Galactic confinement 
volume for CR.
The characteristic time $\tau_{e} (E)$ is in general
a combination of the escape and loss times. The exact form of this combination
is model dependent (see \cite{Lipari:2018usj} for a discussion),
one can  however express it schematically in the form:
\begin{equation}
 \tau_e (E) = \begin{cases}
 T_{\rm esc} (E) \propto E^{-\delta} ~~~ & {\rm for}~~ E \lesssim E^* ~~, \\[0.2 cm]
 T_{\rm esc} (E) \otimes T_{\rm loss} (E) \propto E^{-[1 \oplus\delta]} ~~
 ~~~ & {\rm for} ~~ E \gtrsim E^* \\[0.2 cm]
 \end{cases}
\label{eq:taue}
\end{equation}
where $E^*$ is the characteristic energy where
the loss and escape time are equal ($T_{\rm esc} (E^*) = T_{\rm loss} (E^*)$).

Eq.~(\ref{eq:taue}) expresses the fact that
at low energy ($E \lesssim E^*$) 
when the energy losses are negligible, the characteristic
time for $e^\mp$ coincides with the escape time.
At higher energy 
($E \gtrsim E^*$) one has to combine the
escape time ($T_{\rm esc} \propto E^{-\delta}$)
and the loss time ($T_{\rm loss} \propto E^{-1}$).
The exact form of the combination is model  dependent.
For example \cite{Lipari:2018usj} discusses two examples
where the $E$ dependence of $\tau_e(E)$ at high energy
has the forms $\propto E^{-1}$ and $\propto E^{-(1+\delta)/2}$.
Because of this model dependence  the exponent of the
energy dependence  of $\tau_e(E)$  in Eq.~(\ref{eq:taue}) is written
in a formal, general form  $(1\oplus \delta)$.

After the introduction of these generally accepted concepts,
the standard prediction for the $e^+$ and $\overline{p}$ 
spectra is now  based on the following steps: \\[0.1 cm]

\noindent (S1) The source spectra $Q_{e^+} (E)$ and $Q_{\overline{p}}(E)$
are calculated as the convolution of the spectrum of 
 primary protons (and nuclei) with the appropriate
cross sections:
\begin{equation}
 Q_{{e^+} (\overline{p})}^{\rm sec} (E) \propto
 \int dE_0 ~\phi_p (E_0) ~\left . \frac{d\sigma}{dE} (E,E_0)
 \right|_{pp \to {e^+} (\overline{p})} ~.
\label{eq:q-anti}
\end{equation}
Writing this expression we have assumed that
most of the production of antiparticles happens
in interstellar space (and not inside or near the CR accelerators)
and that the spectra of the CR particles have the same
shape in the entire Galaxy).
Eq.~(\ref{eq:q-anti}) implies that at high energy
(far from threshold effects) one has:
\begin{equation}
 Q_{{e^+} (\overline{p})}^{\rm sec} (E) \propto E^{-\gamma_p}
\label{eq:anti-expo}
\end{equation}
with $\gamma_p$ the spectral index of the proton flux.
Also the positron/anti--proton ratio is  determined by the cross
sections and takes  a value of order
\begin{equation}
\frac{Q_{e^+}(E)}{Q_{\overline{p}}(E)}\approx  2
\label{eq:anti-ratio}
\end{equation}
(see discussion in the appendix of  \cite{Lipari:2016vqk}).
\\[0.1 cm]

\noindent (S2) The properties of the escape time $T_{\rm esc} (E)$
(the absolute scale $T_0$  and the exponent $\delta$ that describes
its rigidity dependence) are obtained from the comparison of 
the spectra of secondary (lithium, beryllium and
boron) and primary (carbon and oxygen) nuclei.
Secondary nuclei are created in the fragmentation of
primary CR nuclei
(in reactions such as $\mathrm{C} + p \to \mathrm {B} +\ldots$,
with $p$ a proton at rest).
Schematically \cite{f:secondary-approx} one has:
\begin{equation}
 \frac{\phi_{\rm B} (E_0)}{\phi_{\rm C} (E_0)} \simeq
 \sigma_{p{\rm C} \to {\rm B}} ~ \frac {X(E_0)}{m_p} \simeq 
 \sigma_{p{\rm C} \to {\rm B}} ~ 
\left \langle n_{\rm ism} \right \rangle \, c \; T_{\rm esc} (E_0) ~.
\label{eq:secondary}
\end{equation}
where $X(E_0)$ is the grammage (that is the column density) of material
crossed by a CR particle before escape, and
$\langle n_{\rm ism} \rangle$ is the average density of the interstellar
medium in the CR confinement volume.
In Eq.~(\ref{eq:secondary}) $E_0$ is the energy per nucleon, 
and we have made use of the fact that in nuclear fragmentations
$E_0$ remains in good approximation constant.
From the study of  the fluxes of secondary nuclei one can the   derive
an exponent $\delta \simeq 0.4$--0.5, and a normalization
$T_{\rm esc} (3~{\rm GeV}) \approx 300$~Myr.
\\[0.25 cm]

\noindent (S3)
The estimate of the escape time obtained  in (S2)
can now  be used to derive a  value of the critical energy $E^*$
that is below a few GeV.
The energy dependence of $\tau_e(E)$  for $E \gtrsim 10$~GeV  takes then
the form:
\begin{equation}
\gamma_e = \gamma_p + (1 \oplus \delta) \simeq 3.4 \div 3.7 
\label{eq:gamma_e}
\end{equation}
(where we used $\gamma_p \simeq 2.7$).
This is the prediction in conflict with the observations.

The same analysis  applied to antiprotons (when energy losses are negligible)
yields the result:
\begin{equation}
\gamma_{\overline{p}} = \gamma_p + \delta \simeq 3.1 \div 3.2 ~~.
\label{eq:gamma_pbar}
\end{equation}

The prediction of the  positron (and antiproton)  spectra  described above
while commonly accepted  raises a number of problems: \\
(P0) It requires a new source of positrons ``fine tuned'' to reproduce the data. \\
(P1) It is also in serious tension (if not in open  conflict)  with
the  data on antiprotons.
The $\overline{p}$ spectrum  (shown in Fig.~\ref{fig:fit_all})
has been  measured  by PAMELA \cite{pamela-antiprotons-2010}
and with more precision by AMS \cite{ams-antiprotons-2016}
and for $E \gtrsim 30$~GeV, has a spectral  index 
$\gamma_{\overline{p}} \simeq 2.8$ that is significantly 
harder than the prediction of Eq.~(\ref{eq:gamma_pbar}).
The disagreement is less significant than for positrons,
and some authors have argued that taking into account all uncertainties
it is possible to reconcile the $\overline{p}$ data with the hypothesis
of a secondary production source \cite{Giesen:2015ufa},
however, all predictions (obtained before the
release of the AMS  $\overline{p}$ data)
estimated a spectrum softer than the data. \\
(P2) The measurements of beryllium isotopes
\cite{beryllium} suggest a CR lifetime shorter than the $T_{\rm esc}(E)$  infered
from the studies of secondary nuclei. \\
(P3) The spectra of electrons and positrons do not show
clearly the expected softening feature associated to the critical energy $E^*$
and the transition to the regime where energy losses are important. \\
(P4) The long lifetime estimated for the CR particles implies slow
propagation. High energy  ($E \gtrsim 1$~Tev) electrons
are therefore expected to only arrive from very few (or just one) young,
near source(s), and the spectrum should exhibit signatures associated 
to this (these)  individual source(s).

The critical  element in the ``standard'' prediction of the
the positron flux (and antiproton flux) is the use of the data
on secondary nuclei to  determine the escape time $T_{\rm esc} (E)$.
A different approach  \cite{Lipari:2016vqk,Lipari:2018usj}
to study the problem is to use the data on positrons and antiprotons
(that in the absence of new sources are  entirely of secondary origin)
to determine the CR propagation properties and determine $T_{\rm esc}(E)$.
The attractive  advantage in this approach is that 
the  hypothesis  of a secondary origin  of $e^+$ and  $\overline{p}$
can be tested  comparing the spectra  because
in this case the  two source  spectra are intimately related
since they they depend  on known cross sections [see Eq.~(\ref{eq:q-anti})].

Inspecting  Fig.~\ref{fig:fit_all}  it is  striking
that in the energy range $E \simeq 30$--400~GeV
the spectra of $e^+$ and $\overline{p}$
have approximately the same spectral index, and  are
close to each other in absolute value  with a ratio $e^+/\overline{p} \approx 2$.
These results are mere coincidences in the standard scenario,
where the two spectra have completely independent sources, and are also
distorted by propagation in different way.
On the other  hand, making use of
Eqs.~(\ref{eq:anti-expo}) and~(\ref{eq:anti-ratio}), one can see
that these observations are  consistent with the hypothesis
that $e^+$ and $\overline{p}$ are both generated in the inelastic
interactions of the same population of primary particles, and
that the effects of propagation  generate distortions 
that are approximately equal.
The study of the antiparticles at lower  energy is also consistent
with this  hypothesis, because the observed relative suppression of
$\overline{p}$ with respect to $e^+$ for $E \lesssim 20$~GeV
(also evident in Fig.~\ref{fig:fit_all})
can be understood  as a  the consequence of the fact  that 
the production of low energy antiprotons is suppressed
for simple  kinematical effects 
(at threshold antiprotons are created in the laboratory
frame with kinetic energy $E_{\overline{p}} \simeq 7~m_p$).

The assumption that $e^+$ and $\overline{p}$ are  both generated as
secondaries and have approximately equal  propagation properties
implies that energy loss effects (for positrons) are small,
and therefore  the critical energy $E^*$ is large
($E^* \gtrsim 400$~GeV), and the  escape time is  much longer
that was is estimated  in the standard model, and also  varies
more  slowly  with rigidity \cite{Lipari:2016vqk}   ($\delta \simeq 0.1$--0.2).

In these scenarios  where the positron and antiproton
determine the escape time,  the interpretation  of
the data on secondary nuclei is problematic. 
These difficulties  can be overcome if most of the
grammage inferred  from the data [see Eq.~(\ref{eq:secondary})]
is accumulated near \cite{Cowsik:2013woa} or inside the sources.

A crucial prediction in this scenario is
that the effects of energy losses must become  visible
at sufficiently high energy, and therefore 
the $e^-$ and $e^+$ spectra should both have softening features
around  a  critical  energy  $E^* \gtrsim 400$~GeV
(see discussion below in Sec.~\ref{sec:softening}).

\section{Spectral hardening}
\label{sec:hardening} 
The high precision AMS data
\cite{Aguilar:2019owu,ams-electrons-positrons-2014}
show that the positron flux undergoes a gradual
hardening in the energy interval at $E \simeq 10$--50~GeV.
Using the fit of \cite{Aguilar:2019owu}, the spectral index
decrease slowly from a maximum ($\gamma \simeq 3.02$)
at $E \simeq 11$~GeV to an approximately constant value
of order $\gamma = 2.75$ for $E \gtrsim  50$~GeV.
It is possible to interpret this hardening
as the transition between two distinct components that are  both of
power law form  but have different spectral indices,
but this interpretation is not unique.
The alternative possibility is that the $e^+$ flux is
dominated by a single component, but the source spectrum,
or the propagation effects do not have a simple power law form.

In this discussion, it is important to note that
also the electron spectrum has a spectral hardening with
a structure very similar to what is observed for positrons.
This point has been already discussed in \cite{Lipari:2018usj},
and is illustrated in Fig.~\ref{fig:flux_components},
that shows together the $e^+$ and $e^-$ spectra measured by AMS.
To help in the visualization and comparison of the two
hardening features, the $e^\mp$ spectra are plotted multiplied by different
energy dependent factors.
The positron spectrum is shown in the form
$E^{2.8} \; \phi_{e^+} (E)$ versus $E$,
while the electron flux is shown in the form
$E^{3.15} \; \phi_{e^-}(E)$ versus $E$, rescaled by a constant factor 1/15.
It is apparent that the deviations of the two spectra from
a simple power law form (that is a straight line in a log--log representation)
have very similar forms.
In fact,
in the entire energy interval 10--300 GeV the energy dependence of the
ratio $\phi_{e^+} (E)/\phi_{e^-} (E)$ is reasonably well described
by a simple power law $\propto E^{+0.35}$, with the effects of the two
hardenings canceling each other.

Both spectra can be described as the sum of two components of power law form
according to Eq.~(\ref{eq:ams-fit}).  
In this case the hardening is  centered  at $E_h$
that is the energy where the contributions of the two components are equal
($E_h \simeq E_0 \; (C_1/C_s)^{1/(\gamma_1 - \gamma_s)}$).
Fits to the electron and positron flux using
the two component form  (without high energy suppression)
have been presented in \cite{Lipari:2018usj}, and give very good
description of the data.
The fit to the $e^-$ spectrum,  and the contributions  two components
are shown in Fig.~\ref{fig:flux_components}, and compared  to the
results for positrons (from the AMS fit).  It is rather striking that
the ``crossing'' energies  for the soft and hard
components are approximately equal  ($E_h^{e^+} \approx E_h^{e^-}$), and also
that the steps in spectral  index across the hardening  features
($\Delta \gamma_\pm \simeq \gamma_1^\pm - \gamma_s^\pm$) are also close
to each other.

For electrons the decomposition of the spectrum
into two components is unexpected, its  interpreration  is therefore
problematic, and it is not clear if one should accept the results of the fit
as evidence for the real existence of two distinct $e^-$ sources. 
It should be stressed that the hard components in the $e^-$
and $e^+$ spectra must have completely different origins
because in the case of electrons this term is
one order of magnitude larger and significantly softer
than the corresponding one for positrons. 

An alternative (purely phenomenological) model to
describe the hardenings in the $e^\pm$ spectra
adopted in \cite{Lipari:2018usj}
is a ``break model'', where the spectrum
(before solar modulations) is described by the form:
\begin{equation}
 \phi(E) = K~\left (\frac{E}{E_0} \right)^{-\gamma_1} ~\left [
1 + \left ( \frac{E}{E_h} \right )^{1/w} 
 \right]^{-\Delta \gamma \, w} ~. 
\label{eq:form-break}
\end{equation}
This functional form depends on 5 parameters
\{$K$, $\gamma_1$, $E_h$, $\Delta \gamma$, $w$ \}
 and describes a spectrum that is the combination of
 two power laws with a change in the spectral
 index around energy $E_h$.
 The break can be either a hardening (for $\Delta \gamma < 0$)
 or a softening (for $\Delta \gamma > 0$).
The spectral index that corresponds to Eq.~(\ref{eq:form-break}) takes the form:
\begin{equation}
 \gamma(E)
 = \gamma_1 + \Delta \gamma ~ 
\left [ 1 + (E_h/E)^{1/w} \right]^{-1}
 = \gamma_1 + \frac{\Delta \gamma}{2} ~ 
 ~\left [ 1 + \tanh\left (\frac{\ln(E/E_h)}{2 \, w} \right ) \right]
\label{eq:gamma-break}
\end{equation}
so that $E_h$ is the energy where the spectral index takes the average
value $\gamma_0 + \Delta \gamma/2$, and $w$ is a width parameters
that in good approximation (see more discussion in \cite{Lipari:2017jou})
gives the interval in $\log_{10} E$ where one half of $\Delta \gamma$ develops.
The functional form of Eq.~(\ref{eq:form-break}) includes the
case of the combination of two separate components
discussed above for the special case where the width parameter takes
the value $w = 1/|\Delta \gamma|$.

Using the form of Eq.~(\ref{eq:form-break}) to describe the $e^-$ and
$e^+$ spectra, one obtains that the best fit values
for the three parameters $E_h$, $\Delta \gamma$ and $w$
that describe the second factor in the equation are approximately equal
(taking into account systematic uncertainties,
also associated to the description of solar modulations).
This is what is expected if the hardenings 
were spectral distortions generated by propagation
effects that are common for electrons and positrons.

In summary, the interpretation of the hardening in the
positron spectrum as evidence for the existence of
two components in the spectrum is not unambiguous.
The interpretation is viable, simple and attractive,
but consistency would then suggest that also the electron
spectrum is formed by two components (perhaps
associated to two different classes of accelerators).
In this scenario the similar structure of the
two hardenings (centered at approximately the same energy
and with approximately the same $\Delta \gamma$)
is a mere coincidence with no physical meaning.

The alternative possibility is that the two 
hardenings in the $e^+$ and $e^-$ spectra
are generated by the same physical mechanism during propagation
of the CR particles in interstellar space.
This idea requires one single mechanism to explain
two different phenomena,
however a realistic physical model for this mechanism 
has not yet been constructed.

\section{High energy suppression}
\label{sec:softening} 
In a discussion about the origin of the
suppression  at high energy (below 1~TeV) of the positron spectrum 
it is important to note that  a marked  softening feature
has also  been observed by several detectors
in the all--electron spectrum
[that is the spectrum for the  sum $(e^- + e^+)$]
at an energy of order 1~TeV.
It is obviously important to study the  relation between these
features.

Measurements  of the all--electron spectrum up to a maximum
energy of 1 and 2~TeV have  been obtained by
AMS \cite{ams02-allelectrons-2014} and Fermi \cite{fermi-electrons-2017}.
These measurements  are  consistent with
an unbroken power law.
Other measurements that reach higher energy
have however observed a spectral  suppression.
The first evidence for the existence of this softening
has been obtained by ground based Cherenkov telescopes
that can measure the spectra of high energy $e^\mp$ selecting events 
that are consistent with an electromagnetic shower,
and subtracting the background generated of hadronic
showers generated by CR protons and nuclei.
A  spectral break was first observed by HESS
\cite{Aharonian:2008aa,Aharonian:2009ah})
and then confirmed by MAGIC \cite{BorlaTridon:2011dk}
and VERITAS \cite{Staszak:2015kza,veritas-electrons}. 
The spectral suppression has then been confirmed
by two calorimeters on satellites in near Earth orbit
CALET \cite{calet-electrons-2017,calet-electrons-2018} and DAMPE \cite{dampe-nature}.

\begin{figure}[bt]
\begin{center}
\includegraphics[width=14.0cm]{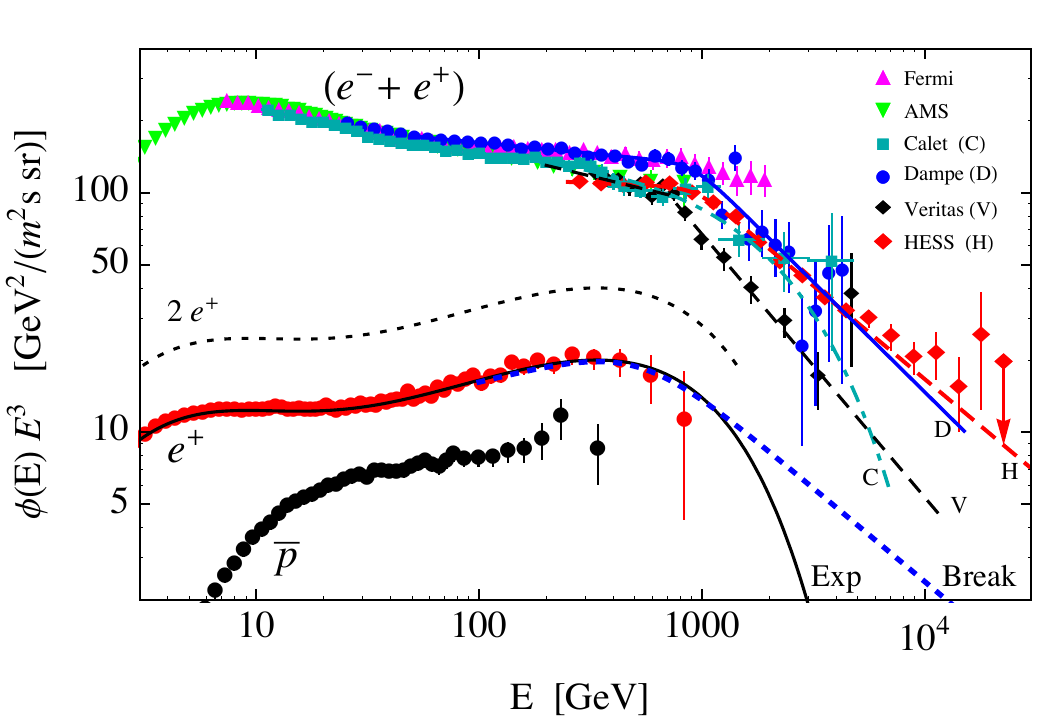}
\end{center}
\caption {\footnotesize
 Spectrum of CR positrons plotted in the form $E^3 \; \phi(E)$.
 The data points are from AMS02 \cite{Aguilar:2019owu}.
 The solid line  (labeled ``Exp'') is the fit  given in the same paper.
 The thick  dashed line (labeled ``Break'')  is a fit to the data
 performed using  the
 functional  form of Eq.~(\ref{eq:form-2breaks}) (see text).
 Also plotted is the AMS spectrum of antiprotons \cite{ams-antiprotons-2016},
 and measurements of the ($e^- + e^+$) spectrum by
 FERMI \protect\cite{fermi-electrons-2017},
 AMS02 \protect\cite{ams02-allelectrons-2014},
 CALET \protect\cite{calet-electrons-2018}, 
 DAMPE \protect\cite{dampe-nature},
 VERITAS \protect\cite{veritas-electrons} 
 and HESS \protect\cite{hess-icrc2017}.
 It should be noted that the VERITAS data  points  do not include
 systematic  uncertainties.
 The lines  labeled (C, D, V and H)
 are the fits  to the ($(e^-+e^+)$ data calculated
 in the papers of DAMPE, CALET, VERITAS, and HESS.
 \label{fig:fit_all}}
\end{figure}

Fig.~\ref{fig:fit_all} show measurements
of the all--electron flux performed by several detectors.
Inspection of the figure shows some discrepancies between the
different data sets (presumably the consequence of
systematic effects), however the existence of a spectral suppression
for $E \gtrsim 1$~TeV is also evident.
All instruments that have obtained measurements in the multi--TeV energy range
observe a suppression of the flux.
The measurements extend  to a maximum energy of order 20~TeV  (in the case
of HESS \cite{hess-icrc2017}), and the  spectral suppression is inconsistent
with an exponential cutoff, but can be   well described as a
``break'' [that is  the functional form of  Eq.~(\ref{eq:form-break})]
with the spectral shapes  below and above $E \simeq 1$~TeV
reasonably well described by power laws with
spectral indices $\gamma_{\rm low} \simeq 3.1$ and
$\gamma_{\rm high} \simeq 3.8$--4.0.

Fig.~\ref{fig:fit_all} shows the best fits curves
estimated (for the their data)  by 
CALET \cite{calet-electrons-2018},
DAMPE \cite{dampe-nature},
VERITAS \cite{veritas-electrons}
and HESS \cite{hess-icrc2017}.
The last  three experiments  use for the fit the spectral break form
\cite{break-parametrization} of Eq.~(\ref{eq:form-break}) 
The fit of the CALET  shown in the figure 
is in the form of an exponential cutoff
($\phi_{(e^- + e^+)} \propto E^{-\gamma} \, e^{-E/E_c}$),
however the Collaboration has also shown
that the data can be equally well fitted  with the break form.

Discussing the relation
between the softening features observed in the $e^+$ and
the $(e^- + e^+)$ spectra a  first important point 
is that in this  energy range the $(e^- + e^+)$ is dominated by electrons
generated in sources (or with mechanisms) different
from those that generate positrons.
To arrive at this conclusion one can observe that for $E \sim 1$~TeV,
the positron flux is only $\sim 15$\% of the all--electron flux.
The sources that generate positrons are expected
to generate also electrons, but it is very likely
that the ratio $e^-/e^+$ at the source is approximately unity
\cite{positron-sources}
(or less for secondary production since proton interactions
generate more $\pi^+$ than $\pi^-$).
The assumption that the positron sources inject equal populations
of $e^+$ and $e^-$ in interstellar space, implies that
these sources account for only approximately
30\% of the all--electron flux
(as illustrated graphically in Fig.~\ref{fig:fit_all}
with the thin dashed line).
The remaining 70\% of the all--electron spectrum at $E \sim 1$~TeV
must therefore be formed by electrons generated in other sources
that emit few (or no) positrons. 
Models such as \cite{Lopez-Coto:2018ksn} where positrons and electrons
are approximately equal for $E \simeq 1$~TeV are in conflict with observations.

The fact that the spectral
break observed in the all--electron spectrum 
corresponds to a feature in the (dominant) component 
of the $e^-$ spectrum that is not related to the positrons.
raises the question of why positrons and electrons
(from a different class of sources) have both
spectral suppressions at a similar energy.

The energy spectrum for a given particle type ($e^+$, $e^-$, $\ldots$)
is determined [as illustrated in Eq.~(\ref{eq:flux-general})]
by the  combination of the source spectrum and of propagation effects
encoded in the characteristic time $\tau_j (E)$.
The possible explanations for the the spectral suppressions observed in the
$e^-$ and $e^+$ fluxes therefore fall into two classes:
\begin{itemize}
\item [(i)] The spectral suppressions in the $e^-$ and $e^+$ fluxes
are associated to the properties of their sources.
However, the sources and/or mechanism that generate the dominant components
in electron and positron fluxes are different,
and therefore the two suppressions must have a different origin.

\item [(ii)] The spectral suppressions are associated to propagation.
The diffusive propagation of $e^\mp$ in interstellar
space is in good approximation independent from the sign of the electric charge
and therefore  propagation effects generate
similar spectral features in the $e^\mp$ spectra.
\end{itemize}

It should be noted hat if the hypothesis (i) is correct, one has the problem
to construct a model to  describe the suppression
of the $e^-$ spectrum for $E \gtrsim 1$~TeV.
An important constraint for such a model is the fact  that 
the suppression is not a cutoff (as  expected if the $e^-$ accelerators
has a sharply defined  maximum energy),  but appears
as a ``break'' with the spectrum  reasonably well described by a power law
in the energy range 1--10~TeV.  On  the other  hand a break--like
softening  structure is what is expected
for a feature generated by energy losses.

On the other hand,
the unambiguous detection of an exponential (or sharper) cutoff
in the positron flux cannot be reconciled
with the hypothesis of a feature generated by  energy losses,
and therefore would exclude hypothesis (ii).
However,  conclusive  evidence for such a sharp cutoff is still missing.
This is illustrated in Fig.~\ref{fig:fit_all}
that shows (as a thick dashed line)  a  fit to the
data  with  using the  expression: 
\begin{equation}
 \phi_{e^+} (E) = K~E^{-\gamma_1}
~\left [
1 + \left ( \frac{E}{E_h} \right )^{1/w_h} 
\right]^{-\Delta \gamma_h \, w_g} 
~\left [
1 + \left ( \frac{E}{E_s} \right )^{1/w_s} 
\right]^{-\Delta \gamma_s \, w_s}
 \label{eq:form-2breaks}
\end{equation}
distorted by solar modulations in the FFA approximation.
The expression in Eq.~(\ref{eq:form-2breaks})  is a power
law  with two  break structures, the first one describes
the hardening at $E \approx 20$~GeV, and  the second one 
the softening below 1~TeV.
The line in Fig.~\ref{fig:fit_all} is calculated with the
parameters $E_s = 520$~GeV, $w_s = 0.35$ and $\Delta \gamma_s =1$
(but a  large  volume in  parameter space yields
fits of  comparable quality), the other parameters
($K$, $\gamma_1$, $\Delta \gamma_h$, $w_h$ and $\varphi$)
are identical  to those  listed in table~I in \cite{Lipari:2018usj}. 
The break form can  provide a  very good fit of the
positron data in the  region of the spectral  suppression.

As discussed in Sec.~\ref{sec:shape} the existence  of softening
features in the $e^+$ and $e^-$ spectra  around the energy  $E^*$
where energy losses  become important 
is an unavoidable prediction for all CR propagation models.
In  the ``standard'' models  for CR  propagation
(that require a new source of positrons)
$E^*$ is of order  few GeV or less, and thefore solutions of class (ii)
for  spectral  features  at $E \sim 1$~TeV are impossible.
It should however also be noted  that it  the imprint of energy loss
should then be present at lower energy,  and it is not clear where,
because both the $e^-$ and $e^+$ spectra below 20~GeV can be modeled
\cite{Lipari:2018usj,Aguilar:2019owu}  as  unbroken power laws
only distorted by solar modulations.

Models where  positrons are of secondary origin put
$E^*$  at an energy of  order one TeV, consistent
with the energy of the observed  spectral sppressions.
The question is if the  shapes of the
spectral  features observed in the $e^+$ and ($e^-+e^+$) spectra
are consistent with the hypothesis  that they are
generated by energy loss effects.

Predictions for these energy loss effects 
have been extensively discussed in \cite{Lipari:2018usj}.
The spectral features generated by the transition to a regime
where energy loss are important
are qualitatively described
[see  Eq.~(\ref{eq:taue})]
by a ``break'',  however their  detailed shape  is model  dependent.
The step in spectral index  takes a value 
$\Delta \gamma \simeq (1 \oplus \delta)-\delta$
that is  between $(1-\delta)/2$ and $(1-\delta)$, while for the width
$w$, the models discussed in \cite{Lipari:2018usj})
give values of order  0.3--1, indicating a rather gradual transition. 

There are indications that
the suppression in the  positron spectrum reported by AMS
is centered at lower energy and is  more  gradual
than the feature  visible in the all--electron flux.
The statistical and systematic  errors are however large,
and there are also  theoretical uncertainties
associated to the fact that the features
imprinted in the $e^-$ and $e^+$ spectra
by energy loss effects are not predicted to be identical. 
Differences can develop for two reasons. The first one is that the
source spectra that are distorted by the propagation effects
have different shapes. The second one is that the space distributions
of the $e^\pm$  sources (that belong to different classes) can be different.
For this reason at this moment it is not possible
to firmly exclude the possibility that the spectral suppressions
are generated by  energy losses during propagation.

\section{Conclusions}
\label{sec:conclusion} 
The interpretation of the CR positron flux is a problem of great
importance for high energy astrophysics.
The first crucial crossroad  in the solution of
this  problem is to determine if the bulk of
high energy positrons can be generated by the
conventional mechanism of secondary production, or
if on the contrary it is necessary
to  have a new source of relativistic positrons.
The first hypothesis requires a revision of some commonly accepted
assumptions about CR propagation in the Galaxy with profound
consequences for our understanding of high energy phenomena.
On the other hand the result of a new positron source
would evidently be a discovery of major importance.

Two features in the positron spectrum can play an important role
in solving the problem of the origin of the positron flux.
The first one is an hardening of the spectrum around $E \approx 25$~GeV
that could be the transition from a low energy regime dominated
by positrons created by the conventional mechanism to a
high energy regime where most of the particles are generated by the new source.
The second one is a suppression at high energy that starts to be visible
for $E$ of order few hundred GeV, and that could mark the maximum energy
for $e^+$ created by the new source.

A comparison  of the positron data  with the spectra of electrons
suggest an  alternative  interpretation for the   feaures
in the $e^+$ spectrum.

The existence of an hardenings with very similar structure
in  $e^-$ spectrum, in the absence of
a convincing two component model for the electron sources,
suggests that  scenarios  where only the hardening
of the positrons is taken as evidence for
the existence of two spectral components could be incorrect.

The $e^+$ and $e^-$ spectra also
have both softening features around an energy of order  $E \sim 1$~TeV.
In this  energy range the dominant components for  the two  spectra
must have their origin in different classes of sources,
and therefore one does not expect that the
two source spectra have  features at the same energy.

On the contrary, models where the bulk of the positrons 
is of secondary origin predict spectral breaks
generated by propagation  that have similar structure
in both the $e^+$ and $e^-$ spectra.
This scenario is however in tension with the
indications that the positron spectral  suppression
develops  at lower energy and is  broader that the
suppression for electrons.

To clarify the situation it is clearly very desirable
to  have more data on the electron and positron spectra
in the TeV and multi--TeV energy range.
The extension to higher energy
of the measurements (with magnetic detectors in space)
of the separate $e^+$ and $e^-$ spectra   are obviously
of great interest.
If this is not possible also new observations
with higher precision, better controlled systematic uncertainties,
and broader converage  at high energy 
of the $(e^+ + e^-)$ spectrum
would give very valuable information.

These experimental studies could
confirm  or refute the  hypothesis of a new positron source,
and doing so also obtain crucially important information
to clarify the physical mechanisms that generate
the $e^+$ and $e^-$ spectra.

\vspace{0.25 cm}
\noindent{\bf Acknowledgments.}
Im grateful to 
Pasquale Blasi,
Carmelo Evoli,
Stefano Gabici,
Philipp Mertsch and Silvia Vernetto
for discussions and exchange of ideas.


\begin{thebibliography}{100}
\bibitem{f:spectral-index}
In this paper
we use the convention used by PAMELA for the definition of the
spectral index: $\gamma (E) = - d\ln\phi/d\ln E$ that has the opposite sign with
respect to the convention used by AMS.
\bibitem{f:interactions}
 A third subdominant sink for CR particles
 is associated to hadronic interactions and  will be neglected here
 for simplicity.
\bibitem{f:rigidity} For simplicity in the notation
 we identify the absolute value of the rigidity and the energy
\bibitem{f:electrons-positrons}
The probability of annihilation is negligible for
relativistic positrons traveling in the interstellar medium.
The energy loss and propagation properties of
relativistic $e^\pm$ in interstellar space are in good approximation equal.
\bibitem{f:secondary-approx}
Equation (\ref{eq:secondary}) is valid  for $X/\lambda_{\rm C} \ll 1$.
\bibitem{break-parametrization}
The DAMPE Collaboration in \cite{dampe-nature} writes the expression
of Eq.~(\ref{eq:form-break}) in terms of the parameter
$s = \Delta \gamma \, w$. 
The choice of $w$ is preferable because it is directly related
to the width of the break [see Eq.~(\ref{eq:gamma-break})] independently
from the value of $\Delta \gamma$.
\bibitem{positron-sources}
 Equal source spectra of $e^-$ and $e^+$ are expected both in
 dark matter models and in positron acceleration models
 when the particles that enter the acceleration mechanism
 are created in $e^+e^-$ pairs.
\end{thebibliography}
\end{document}